\documentclass[a4paper,12pt,dvips,oneside]{article}
\usepackage{rotating,graphicx,overcite,xspace,color,amsmath,amssymb}
\usepackage[scanall]{psfrag}
\usepackage[small,sc]{caption}
\usepackage{amsmath}

\newcommand{\fsz}{\footnotesize}

\newcommand{\mT}{\mathcal{T}}
\newcommand{\mW}{\mathcal{W}}
\newcommand{\mWt}{\widetilde{\mathcal{W}}}
\newcommand{\ds}{\displaystyle}
\newcommand{\dsum}{\displaystyle\sum}

\newcommand{\nuLJ}{\ensuremath{\nu_{\rm LJ}}}

\newcommand{\LJ}[1]{LJ$_{#1}$}
\newcommand{\Pa}{\ensuremath{P_{\rm a}}}
\newcommand{\Pb}{\ensuremath{P_{\rm b}}}
\newcommand{\PA}{\ensuremath{P_{\rm A}}}
\newcommand{\PB}{\ensuremath{P_{\rm B}}}

\begin{document}

\title{Kinetic Analysis of Discrete Path Sampling Stationary Point Databases}
\author{Semen A. Trygubenko\footnote{E-mail: sat39@cam.ac.uk}~~and
David J. Wales\footnote{E-mail: dw34@cam.ac.uk} \\ 
{\it University Chemical Laboratories,
Lensfield Road,} \\ {\it Cambridge CB2 1EW, UK}}
\maketitle
\begin{abstract}
Analysing stationary point databases to extract phenomenological rate constants can become time-consuming for systems with large potential energy barriers. In the present contribution we analyse several different approaches to this problem. First, we show how the original rate constant prescription within the discrete path sampling approach can be rewritten in terms of committor probabilities. Two alternative formulations are then derived in which the steady-state assumption for intervening minima is removed, providing both a more accurate kinetic analysis, and a measure of whether a two-state description is appropriate. The first approach involves running additional short kinetic Monte Carlo (KMC) trajectories, which are used to calculate waiting times. Here we introduce `leapfrog' moves to second-neighbour minima, which prevent the KMC trajectory oscillating between structures separated by low barriers. In the second approach we successively remove minima from the intervening set, renormalising the branching probabilities and waiting times to preserve the mean first-passage times of interest. Regrouping the local minima appropriately is also shown to speed up the kinetic analysis dramatically at low temperatures. Applications are described where rates are extracted for databases containing tens of thousands of stationary points, with effective barriers that are several hundred times $k_BT$.
\end{abstract}

\section{Introduction}
\label{sect:intro}

A great deal of effort is currently focused on the development
of new methods to treat `rare events'.
A number of these methods
employ a coarse-graining approximation of some kind to the phase space
\cite{vanErpMB03,DellagoBG02,BolhuisCDG02,SinghalSP04,FaradjianE04,SriramanKH05,KudinC05}.
Examples include the interface formulation \cite{vanErpMB03}
of transition path sampling \cite{DellagoBG02,BolhuisCDG02},
Markovian state models \cite{SinghalSP04},
milestoning \cite{FaradjianE04},
master equation approaches \cite{BerezhkovskiiS04,SriramanKH05},
and discretized reaction paths \cite{KudinC05}.
The present contribution focuses on the discrete path sampling (DPS)
approach \cite{Wales02,Wales03,Wales04},
which produces a database of stationary points from the underlying
potential energy surface (PES).
Several new methods are developed for extracting phenomenological
two-state rate constants from this database.
As well as speeding up the kinetic analysis, we can also
determine the extent to which a two-state description is appropriate.

The particular coarse-graining approach considered here
focuses on a formally exact partitioning of the PES into
the basins of attraction \cite{mezey81b} of all the local minima \cite{Wales03}.
The corresponding superposition approach to thermodynamics is based
upon the same division of the partition function \cite{stillingerw84,Wales03}.
Kunz and Berry \cite{kunzb95} extended this scheme to treat kinetics by including
transition states, which are here defined geometrically,
as stationary points that possess a single negative Hessian eigenvalue \cite{murrelll68}.
They employed statistical rate theory to calculate individual minimum-to-minimum
rate constants, and a master equation approach \cite{kampen81,kunz95}
to extract global dynamics.
Unfortunately, the number of stationary points is generally expected to scale
exponentially with system size \cite{stillingerw84,WalesD03},
and we must therefore derive an appropriate sampling scheme in order to
represent the kinetics properly.
The DPS approach
was introduced specifically to deal with this problem \cite{Wales02,Wales03}.

To start a DPS analysis we must have an order parameter that
enables us to identify local minima, a$\in$A and b$\in$B, 
which belong to the two end-point states of interest, A and B.
Minima that do not belong to either set will be labelled i$\in$I, where `I' 
stands for `intervening'.
Local equilibrium must be established within each of the A and B regions
relative to their interconversion rate for two-state kinetics to apply.
The original DPS formulation then showed how phenomenological two-state rate constants
$k_{\rm AB}$ and $k_{\rm BA}$ could be written as a sum over discrete paths between A and B,
so long as minima in the I set could be treated according to the steady-state 
approximation \cite{Wales02}.
Here a discrete path is defined as a sequence of local minima and the transition states
that connect them.
Starting from an initial discrete path that connects A and B, 
the DPS procedure generates new discrete paths by
a directed search for additional transition states and local minima that can be
merged with existing paths.
For simple cases, where only a few discrete paths make a significant contribution,
$k_{\rm AB}$ and $k_{\rm BA}$ are obtained from summing the corresponding terms \cite{Wales02}.
However, for more complicated pathways there may be many terms that contribute.
In particular, the databases of stationary points that are constructed during 
the search for new paths generally contain vastly more A$\leftrightarrow$B 
discrete paths than are explicitly considered during the sampling procedure.
However, all these contributions can be summed using a matrix multiplication
approach based upon results from graph theory \cite{Wales02}.
In fact these sums are equivalent to the committor probabilities discussed in
\S \ref{sec:twostate}, and it is more efficient to calculate them using
the successive overrelaxation technique \cite{PressFTV92}.

Overall rates can also be extracted from DPS stationary point databases
using master equation \cite{kampen81} and 
kinetic Monte Carlo \cite{BortzKL75,Voter86,FichthornW91} (KMC) methods \cite{Wales02}.
These approaches do not require the steady-state approximation for minima in the I
set, and hence they enable us to check whether this approximation is valid.
Unfortunately, numerical problems often arise in the master equation approach \cite{Wales02},
although grouping and pruning the stationary point database may help 
to circumvent this issue \cite{Wales03,EvansW03}.
KMC calculations do not experience the same difficulties, but can become
very slow if the potential energy barrier between the A and B regions is large
compared to $k_BT$ \cite{CaiKK02,deKoningCSOKB05}.
Importance sampling KMC schemes have recently been suggested to overcome 
these problems \cite{CaiKK02,deKoningCSOKB05}, but we have not succeeded in
applying them to databases of the complexity considered in the present work. 
Instead we have developed the combined committor probability/KMC 
and graph transformation schemes described 
in the following sections.

\section{Two-State Rate Constant Expressions}
\label{sec:twostate}

We start from a linear master equation formulation of the kinetics \cite{kampen81,kunz95}, 
which immediately invokes the assumption of Markovian dynamics:
\begin{equation}
\begin{array}{lll}
\dfrac{d P_{\rm a}(t)}{dt} &=& \dsum_\alpha k_{\rm a\alpha}\,P_\alpha(t) -  
                            P_{\rm a}(t) \sum_\alpha k_{\alpha \rm a}, \\
\dfrac{d P_{\rm b}(t)}{dt} &=& \dsum_\alpha k_{\rm b\alpha}\,P_\alpha(t) -  
                            P_{\rm b}(t) \sum_\alpha k_{\alpha \rm b}, \\
\dfrac{d P_{\rm i}(t)}{dt} &=& \dsum_\alpha k_{\rm i\alpha}\,P_\alpha(t) -  
                            P_{\rm i}(t) \sum_\alpha k_{\alpha \rm i}, 
\end{array}
\end{equation}
where $P_{\rm \alpha}(t)$ is the occupation probability of minimum $\alpha$ at time $t$
and $k_{\alpha\beta}$ is the rate constant for $\alpha\leftarrow\beta$ transitions between minima
$\alpha$ and $\beta$, which are directly connected by a transition state on the PES.
We assume that minima within the A and B sets are in local mutual equilibrium, so that 
\begin{equation}
\Pa(t)=\frac{\textstyle \Pa^{\rm eq}\PA(t)}{\textstyle \PA^{\rm eq}}  \qquad {\rm and} \qquad 
\Pb(t)=\frac{\textstyle \Pb^{\rm eq}\PB(t)}{\textstyle \PB^{\rm eq}},
\end{equation}
where $P_{\rm A}(t)=\sum_{\rm a\in \rm A}P_{\rm a}(t)$,
etc., and the superscript `eq' stands for `equilibrium'.
In the original derivation of two-state phenomenological rate constants within the DPS
approach we also applied the steady-state approximation for each intervening minimum, i.
These approximations enable us to replace the probabilities $P_{\rm i}(t)$ in an iterative fashion
to obtain
\begin{equation}
k_{\rm AB}^{\rm SS} = \frac{1}{\ds P_{\rm B}^{\rm eq}}  
         \sum_{\rm A\leftarrow B}' \frac{\ds k_{\rm ai_1}\,k_{\rm i_1i_2}\cdots 
          k_{{\rm i}_{n-1}{\rm i}_n}\, k_{{\rm i}_n{\rm b}}\, P_{\rm b}^{\rm eq}}
          {\ds \sum_{\alpha_1} k_{\alpha_1 \rm i_1} \sum_{\alpha_2} k_{\alpha_2 \rm i_2}
          \cdots \sum_{\alpha_n} k_{\alpha_n {\rm i}_n} },
\end{equation}
and an analogous expression for $k^{\rm SS}_{\rm BA}$.
Here the sum is over all discrete paths that begin at any minimum b$\in$B and end
at any minimum a$\in$A, and the prime denotes a restriction that only 
intervening minima i$\in$I can be revisited.
The `SS' superscript is introduced to distinguish this result, which includes the
steady-state approximation for minima in the I set.
Each factor of the form
$k_{\rm j i}/\sum_{\alpha} k_{\alpha \rm i}$ can be written in terms
of the corresponding branching probability, $P_{\rm ji}=k_{\rm ji}/\sum_\gamma k_{\gamma \rm i}$,
where the sum in the denominator runs over all minima $\gamma$ directly connected to $i$:
\begin{equation}
\begin{array}{lll}
k_{\rm AB}^{\rm SS} &=& \dfrac{1}{\ds P_{\rm B}^{\rm eq}}  
         \dsum_{\rm A\leftarrow B}' \ds P_{\rm ai_1}\,P_{\rm i_1i_2}\cdots 
          P_{{\rm i}_{n-1}{\rm i}_n}\, k_{{\rm i}_n{\rm b}}\, \Pb^{\rm eq}, \\
           &=& \dfrac{1}{\ds P_{\rm B}^{\rm eq}}  
         \dsum_{\rm A\leftarrow B}' \ds P_{\rm ai_1}\,P_{\rm i_1i_2}\cdots 
          P_{{\rm i}_{n-1}{\rm i}_n}\, P_{{\rm i}_n{\rm b}}\, \tau_{\rm b}^{-1}\, \Pb^{\rm eq},
\end{array}
\end{equation}
where $\tau_{\rm b}=1/\sum_{\alpha} k_{\alpha \rm b }$ is the mean waiting time
for escape from minimum b to any of its neighbouring minima that are linked to
it by a single transition state.
This expression can be written in a much simpler form by recognising that the
sum of the products of branching probabilities over all paths from minimum b to
minima in the A region that revisit only I minima is the committor probability, $C_{\rm b}^{\rm A}$:
\begin{equation}
	C^{\rm A}_{\rm b} = \dsum_{\rm A \leftarrow \rm b}' P_{\rm ai_1}\,P_{\rm i_1i_2}\cdots
	P_{{\rm i}_{n-1}{\rm i}_n}\, P_{{\rm i}_n{\rm b}},
\end{equation}
so that
\begin{equation}
\label{eq:SS}
k_{\rm AB}^{\rm SS} = \frac{1}{\ds \PB^{\rm eq}}  
         \sum_{\rm b\in \rm B} \frac{\ds C^{\rm A}_{\rm b} \Pb^{\rm eq}}
                           {\ds \tau_{\rm b}}, \qquad {\rm and}
\qquad 
k_{\rm BA}^{\rm SS} = \frac{1}{\ds \PA^{\rm eq}}  
         \sum_{\rm a\in \rm A} \frac{\ds C^{\rm B}_{\rm a} \Pa^{\rm eq}}
                           {\ds \tau_{\rm a}}.
\end{equation}
For example, $C^{\rm A}_{\rm b}$ is the probability that a stochastic trajectory
started from minimum b will encounter an A minimum before a B minimum,
while $C^{\rm B}_{\rm b}=1-C^{\rm A}_{\rm b}$ is the probability that a trajectory will
encounter the B region before A.
The parameter $P^{\rm fold}_\alpha$, defined as the 
probability that a protein will fold before unfolding, starting from some
initial condition $\alpha$ \cite{DuPGTS98,SinghalSP04,SnowSRP04}, 
is a more specific example of a committor probability.

To derive corresponding expressions without invoking the steady-state approximation
for minima in the I set we write the master equation in terms of
transitions between members of the A and B sets and the corresponding effective rate
constants $K_{\rm ab}$ and $K_{\rm ba}$:
\begin{equation}
	\begin{array}{rll}
	k^{\rm NSS}_{\rm AB}
	&=& \ds \frac{1}{\ds \PB^{\rm eq}} \sum_{\rm b\in \rm B}\sum_{\rm a \in A} K_{\rm ab} \Pb^{\rm eq}
	 = \frac{1}{\ds \PB^{\rm eq}}  \sum_{\rm b\in \rm B} K_{\rm Ab} \Pb^{\rm eq}, \\
	\noalign{\medskip}
	{\rm and} \qquad
	k^{\rm NSS}_{\rm BA}
	&=& \ds \frac{1}{\ds \PA^{\rm eq}} \sum_{\rm a\in \rm A}\sum_{\rm b \in B} K_{\rm ba} \Pa^{\rm eq}
	 = \frac{1}{\ds \PA^{\rm eq}}  \sum_{\rm a\in \rm A} K_{\rm Ba} \Pa^{\rm eq},
	\end{array}
\end{equation}
where the superscript `NSS' stands for `non-steady-state'.
$K_{\rm Ab}$ is the effective rate constant for transitions from minimum
b to the A minima, and $K_{\rm Ba}$ is the corresponding rate constant for transitions from minimum
a to the B minima;
the minima in question will not generally be connected by a single transition state.
Treating transitions from minimum b to the A and B regions as independent
Poisson processes with the above rate constants yields a mean
waiting time between transitions of $t_{\rm b} = 1/(K_{\rm Ab}+K_{\rm Bb})$.
Here $K_{\rm Bb}$ corresponds to the effective rate constant for a trajectory
to return to any member of the B set starting from b.
However, $K_{\rm Ab}/(K_{\rm Ab}+K_{\rm Bb})=K_{\rm Ab}t_{\rm b}$ 
can be identified with
the committor probability $C^{\rm A}_{\rm b}$, and hence we obtain
\begin{equation}
\label{eq:committor}
k^{\rm NSS}_{\rm AB} = \frac{1}{\ds \PB^{\rm eq}}  
         \sum_{\rm b\in \rm B} \frac{\ds C^{\rm A}_{\rm b} \Pb^{\rm eq}}
                           {\ds t_{\rm b}}, \qquad {\rm and}
\qquad 
k^{\rm NSS}_{\rm BA} = \frac{1}{\ds \PA^{\rm eq}}  
         \sum_{\rm a\in \rm A} \frac{\ds C^{\rm B}_{\rm a} \Pa^{\rm eq}}
                           {\ds t_{\rm a}}.
\end{equation}
The only difference between the expressions in equations (\ref{eq:SS}) and
(\ref{eq:committor}) is that the average waiting times between events are
interpreted differently.
In (\ref{eq:committor}) $t_{\rm b}$ is the mean waiting time between events
corresponding to transitions between b and any minima in A$\cup$B,
including return to b itself.
However, in (\ref{eq:SS}) $\tau_{\rm b}$ is simply the mean waiting time for
a transition from minimum b to a minimum connected to it by a single transition
state.
Clearly $\tau_{\rm b} \le t_{\rm b}$. 
Furthermore, in the steady-state limit for the intervening minima, the corresponding
waiting times $\tau_{\rm i}$ must be negligible, so that $t_{\rm b} \rightarrow \tau_{\rm b}$,
and $k^{\rm NSS}_{\rm AB}\rightarrow k_{\rm AB}^{\rm SS}$, as expected.

To determine the committor probabilities $C^{\rm A}_{\rm b}$ we use the relation
(a first-step analysis \cite{taylork84})
\begin{equation}
\label{eq:Pfold}
C^{\rm A}_{\rm \alpha} = \sum_{\beta} C^{\rm A}_{\rm \beta} P_{\beta\alpha},
\end{equation}
which is analogous to the expression used for $P^{\rm fold}_\alpha$ in
reference [\citen{SinghalSP04}].
When calculating $k^{\rm SS}_{\rm BA}$ and $k^{\rm NSS}_{\rm BA}$ we
simply replace the `A' superscripts by `B' in equation (\ref{eq:Pfold}).
The sum over $\beta$ includes all minima directly connected to minimum
$\alpha$ by a single transition state, as these
are the only non-zero branching probabilities.
The $C^{\rm A}_{\rm b}$ were calculated iteratively using the successive
overrelaxation technique with an extrapolation factor of 1.999 \cite{PressFTV92}.
The sparse matrix whose elements are the branching probabilities
$P_{\beta\alpha}$
was represented using the compressed row storage scheme \cite{CRS}.

As in previous DPS calculations we can use harmonic densities of states 
to calculate the equilibrium occupation probabilities, along with 
expressions from statistical rate theory \cite{Laidler87} for the individual rate constants,
$k_{\alpha\beta}$,
corresponding to transitions between directly connected minima.
Once the committor probabilities have been calculated we have all the
quantities required to evaluate $k_{\rm AB}^{\rm SS}$ and $k_{\rm BA}^{\rm SS}$.
To calculate the mean waiting time $t_{\rm b}$ for transitions from $\rm b \in B$ 
to any A or B minimum, including revisits to I minima,
we employ the kinetic Monte Carlo (KMC) approach \cite{BortzKL75,Voter86,FichthornW91}.
Stochastic trajectories are simply run from minimum b until we reach any 
minimum belonging to the A or B sets, and $t_{\rm b}$ is evaluated as an 
average over a number of independent trials.

The above formulation in terms of the waiting times $t_{\rm b}$ may have significant
advantages over the alternative KMC approach, where stochastic trajectories 
are followed from minimum b until they reach an A minimum.
This methodology would provide the mean first-passage time from b to A,
$T_{\rm b}$, from which we could calculate the overall rate constants as
\begin{equation}
\label{eq:KMC}
k^{\rm KMC}_{\rm AB} = \frac{1}{\ds P_{\rm B}^{\rm eq}}
 \sum_{\rm b\in \rm B} \frac{\ds P_{\rm b}^{\rm eq}}
                            {\ds T_{\rm b}} \qquad {\rm and} \qquad
k^{\rm KMC}_{\rm BA} = \frac{1}{\ds P_{\rm A}^{\rm eq}}
 \sum_{\rm a\in \rm A} \frac{\ds P_{\rm a}^{\rm eq}}
                            {\ds T_{\rm a}} .
\end{equation}
However, if there is a large potential energy barrier between the A and
B regions then the KMC trajectory will revisit minima in the B region many
times before finally reaching an A minimum. 
$T_{\rm b}$ will then be dominated by a vast number of unsuccessful $b'\leftarrow b$ transitions
\cite{CaiKK02}.
In contrast, the KMC trajectories employed to calculate $t_{\rm b}$
terminate when they hit either an A or a B minimum, and are generally
very short when there is a large barrier.
The vast majority of KMC trajectories return to the B region after a small
number of steps in this situation.

The rate constant formulations that employ committor probabilities in (\ref{eq:SS}) and
(\ref{eq:committor}) will be more efficient that the expressions that use
the mean first-passage time
in (\ref{eq:KMC}) if we can calculate the $C^{\rm A}_{\rm b}$ or
$C^{\rm B}_{\rm a}$ faster than $T_{\rm b}$ or $T_{\rm a}$.
Our experience suggests that this is indeed the case, and that the
corresponding speedup can be very large.
Once the required committor probabilities are known it is also possible
to compare the results from equations (\ref{eq:SS}) and (\ref{eq:committor})
to provide some measure of whether two-state kinetics are appropriate.

\section{`Leapfrog' Moves in Kinetic Monte Carlo}
\label{sec:leapfrog}

At low temperatures we have found that
the KMC runs used to calculate $T_{\rm b}$ and $t_{\rm b}$ may involve an inordinate number
of recrossings for pairs of I minima separated by very low barriers \cite{Wales02}.
This problem was addressed in previous work by evaluating the escape probability
and associated waiting time analytically for such pairs \cite{Wales02}.
A more general scheme was employed in the present work
by calculating the probability and waiting times associated with jumps to 
second-neighbours for each local minimum (`leapfrog' moves). 
A further development involves an expansion of the A and B regions to include 
other local minima, which uses a disconnectivity graph analysis \cite{beckerk97,walesmw98,Wales03},
as described in \S \ref{sec:grouping}.

Consider a particular minimum i, with adjacent minima (connected directly by 
a single transition state) labelled by an index j, and second-neighbours 
labelled by index k$\not=$i. For all minima $\rm j \not\in A\cup B$ we must allow
for an arbitrary number of $\rm i\leftrightarrow \rm j$ recrossings before
escape to a second neighbour, and the recrossings can occur in any order.
Recrossings to minima $\rm j \in A\cup B$ are not allowed, but escape to
such minima must be included. Transitions 
$\rm i \rightarrow \rm j\not\in \rm A\cup \rm B \rightarrow \rm k\not=i$
and $\rm i\rightarrow \rm j\in \rm A\cup \rm B$ can therefore be broken down into independent
events, where the first event involves $\rm i\leftrightarrow \rm j$ recrossings
and the second event involves the actual escape. 
The following derivation is not applicable if neighbour j has a direct
connection to another minimum, $\rm j'$,  that is also adjacent to i, since then
we would need to account for arbitrary $\rm j\leftrightarrow j'$ recrossings as 
well. A more general scheme, which includes such effects, is described
in \S \ref{sec:GT}.

We now define new second neighbour branching probabilities,
$P_{\rm ki}'$, which include all possible recrossings between
i and $\rm j\not\in A\cup B$.
Consider paths from i to k that include $n_{\rm j}$ 
recrossings $\rm i\leftrightarrow \rm j$
for each $\rm j\not\in A\cup B$, with $n=\sum_{\rm j} n_{\rm j}$.
The number of distinct paths of this kind is $n!/\prod_{\rm j} n_{\rm j}!$,
and so the total probability of reaching k from i is
\begin{equation}
\begin{array}{lll}
P_{\rm ki}' &=& \left(\dsum_{\rm j\not\in \rm A\cup \rm B} P_{\rm kj}P_{\rm ji}\right)
\dsum_{n=0}^\infty
\dsum'_{n_{\rm j}} \Bigg(
n!\prod_{\rm j\not\in \rm A\cup \rm B} (P_{\rm ij}P_{\rm ji})^{n_{\rm j}}\Big/
\prod_{\rm j\not\in \rm A\cup \rm B} n_{\rm j}!\Bigg) \\
\noalign{\medskip}
&=& \left(\dsum_{\rm j\not\in \rm A\cup \rm B} P_{\rm kj}P_{\rm ji} \right)
\dsum_{n=0}^\infty
\Bigg(\dsum_{\rm j\not\in \rm A\cup \rm B}P_{\rm ij}P_{\rm ji}\Bigg)^n \\
\noalign{\medskip}
&=& \dsum_{\rm j\not\in \rm A\cup \rm B} P_{\rm kj}P_{\rm ji} \Big/
\Bigg(1-\dsum_{\rm j\not\in \rm A\cup \rm B} P_{\rm ij}P_{\rm ji}\Bigg),
\end{array}
\end{equation}
where we have used the multinomial theorem \cite{Goldberg60},
and the prime for the sum over $n_{\rm j}$ indicates a restriction to 
$n_{\rm j}\ge0$ and $\sum_{\rm j} n_{\rm j}=n$, with $\rm j\not\in \rm A\cup \rm B$.
Similarly, the total probability of reaching a minimum
$j\in \rm A \cup \rm B$ from i is
\begin{equation}
P_{\rm ji}' = P_{\rm ji} \Big/
\Bigg(1-\sum_{\rm j\not\in \rm A\cup \rm B} P_{\rm ij}P_{\rm ji}\Bigg).
\end{equation}
It is easily verified that the new branching probabilities out of i sum
to unity, as they should.

The mean waiting time for a transition from i to any second neighbour, k, or
adjacent minimum, $\rm j\in \rm A\cup \rm B$, is written as $\tau_{\rm i}'$.
To calculate $\tau_{\rm i}'$ we note that every step associated with a
branching probability $P_{\alpha\beta}$ adds $\tau_\beta$ to the duration of
a path, on average. If we replace each branching probability $P_{\alpha\beta}$ by
$\widetilde{P}_{\alpha\beta}=P_{\alpha\beta}\exp(\zeta\tau_\beta)$
then the time associated with any path, multiplied by its probability, can be
obtained from 
\begin{equation}
\left[\frac{d}{d\zeta} \widetilde{P}_{\alpha_1\alpha_2}\widetilde{P}_{\alpha_2\alpha_3}
\widetilde{P}_{\alpha_3\alpha_4}\ldots \widetilde{P}_{\alpha_{n-1}\alpha_n}\right]_{\zeta=0}
=P_{\alpha_1\alpha_2}P_{\alpha_2\alpha_3}
P_{\alpha_3\alpha_4}\ldots P_{\alpha_{n-1}\alpha_n}
(\tau_{\alpha_2}+\tau_{\alpha_3}+\ldots+\tau_{\alpha_n}).  \nonumber
\end{equation}
Hence
\begin{equation}
\begin{array}{lll}
\tau_i' &=&
\left[
\dfrac{d}{d\zeta}
\left(
\dsum_{\rm k}\dsum_{\rm j\not\in \rm A\cup \rm B} \widetilde{P}_{\rm kj}\widetilde{P}_{\rm ji} 
+ 
\dsum_{\rm j\in\rm A\cup \rm B} \widetilde{P}_{\rm ji} \right) \Big/
\Bigg(1-\dsum_{\rm j\not\in \rm A\cup \rm B} \widetilde{P}_{\rm ij}\widetilde{P}_{\rm ji}\Bigg)
\right]_{\zeta=0} \\
\noalign{\medskip}
&=&
\Bigg(\tau_{\rm i} + \dsum_{\rm j\not\in \rm A\cup \rm B} \tau_{\rm j}P_{\rm ji}\Bigg)
\Big/\Bigg(1-\dsum_{\rm j\not\in \rm A\cup \rm B} P_{\rm ij}P_{\rm ji}\Bigg).
\end{array}
\end{equation}

In the present work $\tau_{\rm i}'$ and the corresponding leapfrog 
transition probabilities, $P_{\alpha \rm i}'$, 
were calculated before the KMC phase of the calculation for each minimum
with no direct connections between its first neighbours.
Leapfrog moves were used for eligible minima in the I set if any branching probability
$P_{\alpha \rm i}$ exceeded a threshold value.
In practice, thresholds ranging from 0.1 to 0.8 were found to work equally well,
and gave essentially identical results. The corresponding speedup can be very large,
as discussed in \S \ref{sec:results}.

\section{Reclassification of the A, B and I Local Minima}
\label{sec:grouping}

When reanalysing stationary point databases obtained in previous
work \cite{Wales02,Wales04}, situations were still encountered where the
KMC runs became stuck in extended regions consisting of local minima separated by
small barriers.
This observation suggests that minima in the B region (for A$\leftarrow$B rates)
are connected to relatively
low-lying I minima by barriers that are smaller than those for return to a B minimum.
Expanding the B region to include such minima has a negligible effect on the 
two-state rate constants.
In the present work we have defined expansions of the B (and A) regions
using the same superbasin analysis that is employed in constructing 
disconnectivity graphs \cite{beckerk97,walesmw98,Wales03}.
The local minima are partitioned into disjoint sets (superbasins), so that
at least one pathway exists between any two minima of each set
that does not exceed a threshold energy, $E_{\rm th}$. 
In contrast, any path between minima in different sets must include
a transition state that lies above $E_{\rm th}$.
For a given $E_{\rm th}$, all local minima in the same superbasin as a B minimum were
classified as B, and similarly for the A region.
When the threshold, $E_{\rm th}$, is low enough, this procedure results in no reclassifications;
there is also a maximum threshold above which A and B minima would share the same superbasin.
For the databases considered below, the threshold could be varied over quite a wide
range without changing the rate constants by more than about a factor of two.
However, the corresponding CPU time can vary by orders of magnitude, as
described below.

\section{The Graph Transformation Method}
\label{sec:GT}

The main idea of the graph transformation approach is to progressively remove local minima
from the $\rm I$ set
whilst leaving the average properties of interest unchanged for the database that remains.
The theory is an extension of the results used to perform jumps
to second neighbours in previous KMC simulations,\cite{Wales02}
and the leapfrog moves considered in \S \ref{sec:leapfrog},
which themselves share some common ground with
Novotny's `Monte Carlo with absorbing Markov chains' approach.\cite{novotny94}
The present method
extends the approach of Bortz, Kalos and Lebowitz\cite{BortzKL75}
to exclude not only the transitions from the current state into itself but also transitions
involving an adjacent minimum.
For example, suppose we wish to remove minimum $\rm i\in\rm  I$.
Consider KMC trajectories that arrive at minimum $\beta$, which is adjacent to $\rm i$.
We wish to step directly from $\beta$ to any minimum in the set 
$\Gamma$ that is adjacent to $\beta$ or $\rm i$, excluding these two minima themselves.
To ensure that the mean first-passage times between the $\rm A$ and $\rm B$ sets are unchanged
we must define new branching probabilities, $P_{\gamma\beta}'$ from $\beta$ to all $\gamma\in\Gamma$,
along with a new waiting time for escape from $\beta$, $\tau_\beta'$.
Here, $\tau_\beta'$ corresponds to the mean waiting time for escape from
$\beta$ to any $\gamma\in\Gamma$, while the modified branching probabilities
subsume all the possible recrossings involving minimum $\beta$ that could occur
before a transition to a minimum in $\Gamma$.
Hence the new branching probabilities are:
\begin{equation}
\label{eq:renorm}
P_{\gamma\beta}' = (P_{\gamma \rm i}P_{\rm i \beta}+P_{\gamma\beta}) 
\sum_{m=0}^\infty (P_{\beta \rm i}P_{\rm i  \beta})^m =
(P_{\gamma \rm i}P_{\rm i\beta}+P_{\gamma \beta})/(1-P_{\beta \rm i}P_{\rm i \beta}).
\end{equation}
This formula also applies if either $P_{\gamma \beta}$ or $P_{\gamma i\rm }$ vanishes.
When calculating mean first-passage times for transitions from the B to A regions we
do not consider branching probabilities out of A minima, and similarly for 
branching probabilities out of B minima when considering A to B transitions.
However, $P_{\gamma\beta}'$ and $\tau_\beta'$ for $\beta\in \rm A$ are never used in
calculating the $\rm A\leftarrow B$ rates, and similarly for $\beta\in \rm B$
and $\rm B\leftarrow A$ rates.
Hence we can use equation (\ref{eq:renorm}) for all minima $\beta \in \rm A\cup B\cup I$,
and obtain results for both $\rm A\leftarrow \rm B$ and $\rm B\leftarrow \rm A$
rate constants at the end of the procedure. 
Detailed balance can then be used as a consistency check.

It is easy to show that the new branching probabilities are normalised:
\begin{equation}
\sum_{\gamma\in\Gamma} \frac{P_{\gamma \rm i}P_{\rm i \beta}+P_{\gamma \beta}}{1-P_{\beta \rm i}P_{\rm i \beta}}
= \frac{(1-P_{\beta \rm i})P_{\rm i \beta}+(1-P_{\rm i \beta})}{1-P_{\beta \rm i}P_{\rm i \beta}}
= 1.
\end{equation}
To calculate $\tau_{\rm\beta}'$ we use the method of Sec.~\ref{sec:leapfrog}:
\begin{equation}
\tau_\beta' = \left[ \frac{d}{d\zeta}
\sum_{\gamma\in\Gamma}\frac{P_{\gamma \rm i}P_{\rm i \beta}e^{\zeta(\tau_{\rm i}+\tau_\beta)}
+P_{\gamma \beta}e^{\zeta\tau_\beta}}{1-P_{\beta {\rm i}}P_{\rm i \beta}e^{\zeta(\tau_{\rm i}+\tau_\beta)}} \right]_{\zeta=0}
= \frac{\tau_\beta+P_{\rm i \beta}\tau_{\rm i}}{1-P_{\beta \rm i}P_{\rm i \beta}}.
\end{equation}
The modified branching probabilities and waiting times could be used in a KMC
simulation based upon the stationary point database with minimum $\rm i$ excluded.
Since the modified branching probabilities, $P_{\gamma \beta}'$, subsume the sums over all 
paths from $\beta$ to $\gamma$ that involve $\rm i$ we would expect the probability that a
trajectory starting at $\rm b\in \rm B$ and ending at $\rm a\in \rm A$ is conserved.
Here we consider for specificity the $k_{\rm AB}$ rate constant, corresponding to paths
that start from $\rm B$ minima and terminate at an $\rm A$ minimum. 
In contrast to the formulation of 
$k_{\rm AB}^{\rm SS}$ and $k_{\rm AB}^{\rm NSS}$, above, 
revisits to B minima are now allowed, as for the conventional KMC 
calculations that yield $k_{\rm AB}^{\rm KMC}$ in equation (\ref{eq:KMC}).
Since each trajectory exiting from $\gamma\in\Gamma$ acquires a time increment equal to
the average value, $\tau_{\rm\beta}'$, the contributions to the mean first-passage time from individual A minima
are not conserved (unless there is a single A minimum).
Nevertheless, the overall mean first-passage time to $\rm A$ is
conserved, i.e.~$\mT_{\rm b}=\mT_{\rm b}'$,
where the prime denotes a transformed quantity.
To prove these results 
consider the effect of removing minimum i on trajectories reaching minimum $\beta$,
which is initially connected to i, from minimum $\rm b\in\rm B$. 
The total probability of a pathway terminating at a if it starts from b is the
sum of the product of branching probabilities, 
$\mW_\xi=P_{\rm a \xi_n}P_{\xi_n\xi_{n-1}}\ldots P_{\xi_2\xi_1}P_{\xi_1\rm b}$,
over all the corresponding paths $\xi\in \rm a\leftarrow \rm b$:
\begin{equation}
\begin{array}{lll}
\dsum_{\xi\in \rm a\leftarrow \rm b} \mW_\xi 
        &=& \dsum_{\xi_3\in\Xi'} \mW_{\xi_3} + \dsum_{\xi_1 \in \beta \leftarrow b} \mW_{\xi_1}
    \dsum_{\gamma\in\Gamma} (P_{\gamma \beta}
    +P_{\gamma \rm i}P_{\rm i \beta})\dsum_{m=0}^\infty(P_{\beta \rm i}P_{i \beta})^m
    \dsum_{\xi_2\in a\leftarrow\gamma} \mW_{\xi_2} \\
&=& \dsum_{\xi_3\in\Xi'} \mW_{\xi_3} + \dsum_{\xi_1 \in \beta \leftarrow \rm b} 
    \mW_{\xi_1} \dsum_{\gamma\in\Gamma} P_{\gamma \beta}'
    \dsum_{\xi_2\in \rm a\leftarrow\gamma} \mW_{\xi_2},
\end{array}
\end{equation}
and similarly for any other minimum adjacent
to i and any other pathway that revisits $\beta$ other than
by immediate recrossings of the type $\beta\rightarrow {\rm i} \rightarrow\beta$.
$\Xi$ is the ensemble of all paths for which probabilistic weights
cannot be written in the form defined by the last term of the above
equation.
Here $\Xi'$ is the ensemble of all paths
which probabilistic weights cannot be written in the form defined by the last term of the above equation.
Hence the transformation preserves
the individual probabilities $\sum_{\xi\in \rm a\leftarrow \rm b} \mW_\xi$.

Now consider the effect of removing minimum $\rm i$ on the contribution to
the mean first-passage time from $\rm b\in \rm B$ to A
using the approach of Sec.~\ref{sec:leapfrog}:
\begin{equation}
        \begin{array}{lll}
&& \ds \left[\frac{d}{d\zeta} 
    \sum_{\xi_1 \in \rm \beta\leftarrow \rm b} \mWt_{\xi_1} 
    \sum_{\gamma\in\Gamma} \widetilde{P}_{\gamma\beta}'
    \sum_{\xi_2\in a\leftarrow\gamma} \mWt_{\xi_2} \right]_{\zeta=0}  \\
\noalign{\medskip}
&=& \ds \sum_{\xi_1 \in \beta\leftarrow \rm b} 
       \left[ \frac{d\mWt_{\xi_1}}{d\zeta} \right]_{\zeta=0}
    \sum_{\gamma\in\Gamma} {P}_{\gamma\beta}'
    \sum_{\xi_2\in a\leftarrow\gamma} \mW_{\xi_2}  \\
\noalign{\medskip}
&& \ds +
    \sum_{\xi_1 \in \beta\leftarrow \rm b} \mW_{\xi_1} 
    \sum_{\gamma\in\Gamma} \left[ \frac{d\widetilde{P}_{\gamma\beta}'}{d\zeta} \right]_{\zeta=0}\
    \sum_{\xi_2\in a\leftarrow\gamma} \mW_{\xi_2}   \\
\noalign{\medskip}
&& \ds +
    \sum_{\xi_1 \in \beta\leftarrow \rm b} \mW_{\xi_1} 
    \sum_{\gamma\in\Gamma} {P}_{\gamma\beta}' 
    \sum_{\xi_2\in \rm a\leftarrow\gamma} \left[ \frac{d\mWt_{\xi_2}}{d\zeta} \right]_{\zeta=0},
    \end{array}
\end{equation}
where the tildes indicate that every branching probability $P_{\alpha \beta}$ is
replaced by $P_{\alpha \beta}e^{\zeta\tau_\beta}$, as in \S \ref{sec:leapfrog}.
The first and last terms are unchanged from the original database in this construction, 
but the middle term,
\begin{equation}
        \begin{array}{lll}
&& \ds \sum_{\xi_1 \in \beta\leftarrow b} \mW_{\xi_1} 
    \sum_{\gamma\in\Gamma} \left[ \frac{d\widetilde{P}_{\gamma \beta}'}{d\zeta} \right]_{\zeta=0}\
    \sum_{\xi_2\in \rm a\leftarrow\gamma} \mW_{\xi_2}  \\
\noalign{\bigskip}
&=& \ds \sum_{\xi_1 \in \beta\leftarrow b} \mW_{\xi_1}
  \sum_{\gamma\in\Gamma} \frac{P_{\gamma \rm i}P_{\rm i \beta}(\tau_\beta+\tau_{\rm i})
+P_{\gamma \beta}(\tau_\beta+P_{\beta \rm i}P_{i \beta}\tau_{\rm i})}
                            {(1-P_{\beta \rm i}P_{\rm i \beta})^2}
  \sum_{\xi_2\in \rm a\leftarrow\gamma} \mW_{\xi_2},
        \end{array}
\end{equation}
is different (unless there is only one A minimum).
However, if we sum over A minima then 
$\sum_{\rm a\in \rm A}\sum_{\xi_2\in \rm a\leftarrow\gamma} \mWt_{\xi_2}=1$ for
all $\gamma$, and we can now simplify the sum over $\gamma$ as
\begin{equation}
        \sum_{\gamma\in\Gamma}
\frac{P_{\gamma \rm i}P_{\rm i \beta}(\tau_\beta+\tau_{\rm i})+P_{\gamma \beta}
     (\tau_\beta+P_{\beta \rm i}P_{\rm i \beta}\tau_{\rm i})}
     {(1-P_{\beta \rm i}P_{\rm i \beta})^2} = \tau_\beta' = \sum_{\gamma\in\Gamma}
P_{\gamma \beta}'\tau_\beta'.
\end{equation}
The same argument can be applied whenever a trajectory reaches a minimum adjacent to $\rm i$,
so that $\mT_{\rm b}'=\mT_{\rm b}$, as required.

At low temperature some of the $P_{\rm \alpha\beta}$ approach unity, and 
numerical problems arise in calculating terms like $1-P_{\alpha\beta}P_{\beta\alpha}$.
However, precision can be regained by writing 
\begin{equation}
P_{\alpha\beta}=1-\sum_{\gamma\not=\alpha}P_{\gamma\beta}\equiv 1-\epsilon_{\alpha\beta} \qquad
{\rm and} \qquad
P_{\beta\alpha}=1-\sum_{\gamma\not=\beta}P_{\gamma\alpha}\equiv 1-\epsilon_{\beta\alpha}, 
\end{equation}
and then using
$1-P_{\alpha\beta}P_{\beta\alpha}=
\epsilon_{\alpha\beta}-\epsilon_{\alpha\beta}\epsilon_{\beta\alpha}+\epsilon_{\beta\alpha}$.

Further applications of the graph transformation method will be described elsewhere
in the context of Markov chains considered as digraphs.\cite{GT}
In particular, we note that more general formulations are possible so that waiting
times involving revisits to arbitrary sets of minima can be included.
The main advantage of this theory is that there is an upper bound to the maximum
operation count involved in removing all the I minima.
In contrast to the successive overrelaxation technique used to calculate
committor probabilities, described in \S \ref{sec:twostate},
the graph transformation does not depend upon satisfying any convergence criteria.
Once all the I minima have been removed the only transformed branching
probabilities remaining, $P_{\rm a \rm b}'$, are the relative probabilities of paths
from  A and B minima terminating at any of the other A and B minima. 
The final waiting times, $\tau_{\rm b}'$, correspond to the average time before a
transition to any other A or B minimum starting from b, and similarly for the $\tau_{\rm a}'$. 
Here we use a single prime to denote the final values of transformed quantities,
although in general they may change many times as the I minima are successively removed.
The associated rate constants can then be calculated as
\begin{equation}
\label{eq:GT}
k^{\rm GT}_{\rm AB} = \frac{1}{\ds \PB^{\rm eq}}  
         \sum_{\rm b\in \rm B}\frac{\ds \Pb^{\rm eq}}{\ds \tau_{\rm b}'}
                           \sum_{\rm a\in A} P_{\rm ab}'  \qquad {\rm and}
\qquad 
k^{\rm GT}_{\rm BA} = \frac{1}{\ds \PA^{\rm eq}}  
         \sum_{\rm a\in \rm A}\frac{\ds \Pa^{\rm eq}} {\ds \tau_{\rm a}'}
                           \sum_{\rm b\in \rm B} P_{\rm ba}'.
\end{equation}

The transformed probabilities and waiting times in equation (\ref{eq:GT}) are different from the 
committor probabilities and waiting times that appear in the definitions of $k_{\rm AB}^{\rm SS}$
and $k_{\rm AB}^{\rm NSS}$, and from the waiting times that appear in 
equation (\ref{eq:KMC}) for $k_{\rm AB}^{\rm KMC}$.
It is also noteworthy that we have avoided the steady-state approximation for the I minima.
Since the computational cost of the graph transformation does not change
as the temperature decreases, in contrast to the successive overrelaxation 
calculations and conventional KMC runs, this approach becomes increasingly
advantageous at low temperature.
In fact it is also possible to define transformations that enable us to
evaluate $k_{\rm AB}^{\rm NSS}$ and $k_{\rm AB}^{\rm KMC}$ precisely in a 
fixed number of operations; these results will be presented elsewhere.\cite{GT}

\section{Results}
\label{sec:results}

The formulations in equations (\ref{eq:SS}),
(\ref{eq:committor}), and (\ref{eq:GT}) were tested for stationary point databases obtained in previous
DPS runs for permutational isomerization and morphological transitions of 
four different atomic clusters bound by the Lennard-Jones (LJ) potential \cite{jonesi25}.
In each case Arrhenius fits of $k_{\rm AB}$ and $k_{\rm BA}$ were presented in
previous work \cite{Wales02}, and only for the cluster containing 55 atoms, \LJ{55},
are the results significantly different here.
For the three other examples (\LJ{13}, \LJ{38} and \LJ{75})
the effective barrier heights changed by less than
$0.05\,\epsilon$, and the Arrhenius prefactors changed by a factor of three or less.
Here $\epsilon$ is the pair well depth for the LJ potential. 
The new results for \LJ{55} are therefore considered in more detail below.

The global potential energy minimum for \LJ{55} is a Mackay icosahedron \cite{mackay62},
which exhibits special stability and `magic number' properties \cite{berrybdj88,labastiew90}.
There are four distinct sites for a tagged atom in the global minimum:
one in the centre, one in the middle shell, and two in the outer shell \cite{Wales02}.
DPS calculations in previous work considered the rate for migration of the tagged
atom between the centre and either one of the surface sites \cite{Wales03}.
Even at the solid-like/liquid-like melting transition temperature for this cluster the
potential energy barrier between these A and B states is around $50\,k_BT$.
Temperatures down to $k_BT/\epsilon=0.04$ were considered in the present work,
where the barrier is $350\,k_BT$.
Standard KMC calculations for this system previously proved to be
unfeasible below about $k_BT/\epsilon=0.3$ \cite{Wales02}.

A disconnectivity graph that distinguishes permutation-inversion isomers of the tagged atom
is shown in Figure \ref{fig:LJ55.permtree}.
Separate branches appear for the two distinct surface sites, but these minima were
grouped together in state B for the rate constant calculations.
The branch for group A, at the right of the graph, has the tagged atom
at the centre of the icosahedron.
The calculated two-state rate constants do not vary significantly
over a wide range of $E_{\rm th}$,
and KMC averages for $t_{\rm b}$ employing
1000 trajectories are generally sufficient.
For higher thresholds the A and B sets start to overlap, indicating
that the highest energy transition state on the lowest energy path between
the A and B regions lies approximately 14$\,\epsilon$ above the global minimum.
Using a fractional convergence tolerance of $10^{-3}$ for the total rate 
in the committor probability calculations
and KMC averages over 1000 trajectories per B minimum, the entire 
analysis of the database
requires only a few seconds of CPU time on an UltraSparcIII 900\,MHz processor.

The temperature dependence of
both $k^{\rm NSS}_{\rm AB}$ and $k^{\rm NSS}_{\rm BA}$, 
for the surface-to-centre and centre-to-surface rates, 
can be fitted quite accurately (coefficient of determination, $R^2=0.99999$)
by the Arrhenius form $k=a\,\exp(-\Delta/k_BT)$ for $0.04\le k_BT/\epsilon \le 0.3$:
\begin{equation}
\begin{array}{lll}
&& k_{\rm AB}: \qquad a\,=\,1.38\times10^{8}\,\nuLJ , \quad  \Delta=14.05\,\epsilon, \\
&& k_{\rm BA}: \qquad a\,=\,6.52\times10^{9}\,\nuLJ , \quad  \Delta=14.04\,\epsilon, 
\end{array}
\end{equation}
where the unit of frequency is $\nuLJ=\sqrt{\epsilon/M\sigma^2}$, with
$M$ the atomic mass and $2^{1/6}\sigma$ the equilibrium pair separation.
Symmetry requires that $42k_{\rm AB}=k_{\rm BA}$
when only permutation-inversion isomers of the global minimum including a
tagged atom are
included in the A and B sets \cite{Wales02}.
The above results deviate a little from this ideal ratio because there are 77 minima in
the A set and 1763 minima in the B set using reclassification at a threshold
energy of $E_{\rm th}=-270\,\epsilon$.
At $k_BT/\epsilon=0.1$ the centre-to-surface rate constant
has order $10^{-40}$\,s$^{-1}$ for parameters appropriate to argon.
 
The above rates are significantly faster than those obtained from a summation
over DPS paths in previous work \cite{Wales02}, indicating that the latter 
sums were not converged. 
In particular, the value of the effective barrier,
$\Delta$, in the Arrhenius fit is about 4$\,\epsilon$ lower, and now
coincides closely with the barrier corresponding to the highest transition state on the
lowest energy path between the A and B regions.
The present results therefore supersede the values obtained in reference
[\citen{Wales04}].

A more detailed analysis of the results for \LJ{55} (Tables \ref{table:four}-\ref{table:one})
reveals a number of trends that are likely to be useful in future work,
especially regarding the choice of $E_{\rm th}$ and two-state kinetics.
At $k_BT/\epsilon=0.4$ we find that $k^{\rm SS}_{\rm AB}$, $k^{\rm NSS}_{\rm AB}$
and $k^{\rm GT}_{\rm AB}$
agree well for $-267\,\epsilon\ge E_{\rm th} \ge -270\,\epsilon$ (Table \ref{table:four}).
These values also agree with $k^{\rm KMC}_{\rm AB}$ calculated without any
regrouping, although the latter simulation requires about
a thousand times more CPU time. 
It is also interesting to note that the calculated values in Table \ref{table:four}
exhibit jumps when $E_{\rm th}$ changes from $-270\,\epsilon$ to $-271\,\epsilon$,
from $-273\,\epsilon$ to $-274\,\epsilon$, and from $-275\,\epsilon$ to $-276\,\epsilon$.
Changes in behaviour occur at the same points for lower temperatures, as
described below, and the cause is easily identified with the aid of Figure \ref{fig:LJ55.permtree}.
For $E_{\rm th}=-270\,\epsilon$ the Mackay icosahedra corresponding to both
minima in the original B set, and the $C_{5v}$ minimum with the tagged atom in the
intermediate shell, all lie in the same superbasin. 
According to the reclassification scheme, all the local minima belonging
to this superbasin are then classified as type B.
However, at $E_{\rm th}=-271\,\epsilon$ the branch corresponding to the 
tagged atom in the intermediate shell splits off from the branch containing the
original B minima, so there is a qualitative change in the classification of
local minima between these energies.
Nevertheless, the overall A$\leftrightarrow$B kinetics are not disturbed
dramatically, because equilibrium between the remaining B minima and 
the region that is now classed as `intervening' is still established rapidly
compared to the time scale for transitions between A and B.
As Figure \ref{fig:LJ55.permtree} clearly shows, the corresponding potential energy
barrier for A$\leftrightarrow$B interconversion is about 5\,$\epsilon$ 
larger than for equilibration with this I region.
It is also noteworthy that the relation $42k_{\rm AB}=k_{\rm BA}$
is obeyed to better than $0.2\%$ for $E_{\rm th}=-271\,\epsilon$,
because the A and B sets are not expanded as much as for $E_{\rm th}=-270\,\epsilon$,
where the Arrhenius fit was performed (above).

On changing $E_{\rm th}$ from $-273\,\epsilon$ to $-274\,\epsilon$
the two icosahedra with the tagged atom in the surface separate into different
superbasins. 
Some local minima corresponding to icosahedra with a surface vacancy and adatom
pair also separate into their own branches, and are therefore assigned to the I set.
This regrouping again affects $k^{\rm SS}_{\rm AB}$, $k^{\rm NSS}_{\rm AB}$,
and $k^{\rm GT}_{\rm AB}$, but the difference is only a factor of about two.
Nevertheless, the change in $k^{\rm NSS}_{\rm AB}$ below $E_{\rm th}=-270\,\epsilon$
suggests that a two-state description is less appropriate for the classification
of A, B and I sets corresponding to lower energy thresholds.

When the temperature is lowered to $k_BT/\epsilon=0.3$ we again see
changes in the calculated rate constants at the same
$E_{\rm th}$ boundaries as for $k_BT/\epsilon=0.4$.
However, there is one important change from the results at $k_BT/\epsilon=0.4$.
A much tighter convergence condition is required in the calculation
of committor probabilities for $E_{\rm th}\le-271\,\epsilon$ at the lower temperature,
and the required CPU time jumps by a factor of about $100$.
This effect can again be explained from the corresponding classification of the
A, B and I sets with reference to the disconnectivity graph.
The committor probability calculation requires many more iterations to converge
for $E_{\rm th}\le-271\,\epsilon$ because there are I minima with very low
values of $C^{\rm A}_{\rm i}$.
The relative CPU time required to calculate $k^{\rm GT}_{\rm AB}$ 
is now significantly smaller, since it is essentially unchanged from $k_BT/\epsilon=0.4$.

For $k_BT/\epsilon=0.2$ the benefit of leapfrog KMC moves becomes visible (Table
\ref{table:two}), producing a speedup factor of between ten and a few hundred.
An even tighter condition is required to converge the committor probability calculation 
for $E_{\rm th}\le -271\,\epsilon$, with a corresponding dramatic increase in
the time required.
At $k_BT/\epsilon=0.1$ (Table \ref{table:one})
KMC runs without leapfrog moves were not feasible, and only $k^{\rm GT}_{\rm AB}$
could be calculated for $E_{\rm th}\le-271\,\epsilon$.
Note that the timings for $k^{\rm GT}_{\rm AB}$ are practically independent of
temperature, which makes the graph transformation approach highly advantageous
at low temperatures.

\section{Conclusions}
\label{sec:conc}

In this contribution we have shown how the original discrete path sampling (DPS)
rate expressions,
which involve the steady-state approximation for `intervening'
local minima, can be written in terms of committor probabilities.
Analogous expressions can be derived without the steady-state approximation
by simply changing the waiting time associated with escape from each B (for
$k_{\rm AB}$) or A (for $k_{\rm BA}$) minimum.
In the steady-state expressions the appropriate waiting times correspond to transitions to
directly connected minima, involving a single transition state.
In contrast, when the steady-state approximation is removed, the required waiting times
correspond to transitions to any member of the A or B sets, 
and can be calculated from relatively short KMC runs.
If the waiting times for escape from each intervening local minima to an
adjacent minimum are small then the steady-state limit is recovered, as expected.

Committor probabilities can be calculated using successive
overrelaxation techniques, and this approach is significantly faster than
the direct sum over paths of increasing length used in previous work to
calculate $k^{\rm SS}_{\rm AB}$ and $k^{\rm SS}_{\rm BA}$ \cite{Wales02,Wales04}.
The KMC runs required to calculate waiting times for transitions to either the
A or B region, and hence $k^{\rm NSS}_{\rm AB}$ or $k^{\rm NSS}_{\rm BA}$,
are generally very short. 
In contrast, direct KMC simulations based on mean first-passage times involve trajectories
that can only terminate in the product region.
When the A and B states are separated by a high potential energy barrier the
latter trajectories will generally revisit the reactant region many times.
The alternative formulations based upon committor probabilities
may then be advantageous, but require leapfrog KMC moves and reclassification
of the A, B and I sets to be feasible at low temperatures.

The most powerful method that we have found for extracting phenomenological
rate constants is the graph transformation approach. Intervening 
local minima are successively removed, and the branching probabilities
and waiting times for the remaining local minima are renormalised to
preserve the mean first-passage time between the A and B regions.
At the end of the transformation only A and B minima remain, and the
waiting time in each one is the average value for a transition to any
of the other A and B minima.
The advantage of this approach is that the time taken to perform the
transformation does not depend upon temperature, so it can be used
when the processes of interest are arbitrarily slow.

The main conclusion that we draw from the detailed analysis of \LJ{55} in
\S \ref{sec:results} is that the threshold energy, $E_{\rm th}$,
for reclassification of A and B minima should be chosen above
the potential energy where any likely kinetic traps branch off.
This choice of threshold should not affect the two-state dynamics of interest significantly,
so long as the potential energy difference between the top of any possible
traps and the highest transition state on the lowest $\rm A \leftrightarrow \rm B$
path is large compared to $k_BT$.
In this case there will be a clear separation of time scales for relaxation between the
A and B regions and within the regions themselves, so that a two-state description
is still applicable \cite{Chandler78,DellagoBC99}.
The disconnectivity graph approach \cite{beckerk97,walesmw98} provides a helpful way to
recognise such situations \cite{Wales03}.
Both the graph transformation theory and
leapfrog KMC moves combined with an appropriate choice of $E_{\rm th}$ can 
produce exponential speedups in the kinetic analysis, providing access
to temperatures where the potential energy barrier is more than a hundred
times larger than $k_BT$.

\def\cam{Cambridge University Press, Cambridge }
\def\dover{Dover Publications, New York }
\def\mit{The MIT Press, Cambridge, Massachusetts }
\def\wiley{John Wiley and Sons, New York }
\def\springer{Springer-Verlag, New York }
\def\elsevier{Elsevier, Amsterdam }
\def\aciee{Angew. Chem. Int. Ed. Engl. }
\def\ac{Acta. Crystallogr. }
\def\acp{Adv. Chem. Phys. }
\def\acr{Acc. Chem. Res. }
\def\ajp{Am. J.~Phys. }
\def\ap{Ann. Physik }
\def\apc{Adv. Prot. Chem. }
\def\arpc{Ann. Rev. Phys. Chem. }
\def\bi{Bioinf. }
\def\bj{Biophys. J. }
\def\cccc{Coll. Czech. Chem. Comm. }
\def\cpc{Comp. Phys. Comm. }
\def\crev{Chem. Rev. }
\def\el{Europhys. Lett. }
\def\ic{Inorg. Chem. }
\def\ijmpc{Int. J.~Mod. Phys. C }
\def\ijqc{Int. J.~Quant. Chem. }
\def\jcis{J.~Colloid Interface Sci. }
\def\jcsft{J.~Chem. Soc., Faraday Trans. }
\def\jacs{J.~Am. Chem. Soc. }
\def\jas{J.~Atmos. Sci. }
\def\jbc{J.~Biol. Chem. }
\def\jcc{J.~Comp. Chem. }
\def\jcp{J.~Chem. Phys. }
\def\jce{J.~Chem. Ed. }
\def\jcscc{J.~Chem. Soc., Chem. Commun. }
\def\jetp{J.~Exp. Theor. Phys. (Russia) }
\def\jmb{J.~Mol. Biol. }
\def\jmsp{J.~Mol. Spec. }
\def\jmst{J.~Mol. Struct. }
\def\jncs{J.~Non-Cryst. Solids }
\def\jpa{J.~Phys. A }
\def\jpc{J.~Phys. Chem. }
\def\jpca{J.~Phys. Chem. A }
\def\jpcb{J.~Phys. Chem. B }
\def\jpcm{J.~Phys. Condensed Matter. }
\def\jpcs{J.~Phys. Chem. Solids. }
\def\jpsj{J.~Phys. Soc. Jpn. }
\def\jrnist{J.~Res. Natl. Inst. Stand. Technol. }
\def\mg{Math. Gazette }
\def\nat{Nature }
\def\nsb{Nat. Struct. Biol.}
\def\Pa{Physica A }
\def\Pd{Physica D }
\def\pac{Pure. Appl. Chem. }
\def\pccp{Phys. Chem. Chem. Phys. }
\def\phys{Physics }
\def\pmb{Philos. Mag. B }
\def\ptrsa{Philos. T. Roy. Soc. A }
\def\ptrsb{Philos. T. Roy. Soc. B }
\def\pnasu{Proc. Natl. Acad. Sci. USA }
\def\pr{Phys. Rev. }
\def\prep{Phys. Reports }
\def\pra{Phys. Rev. A }
\def\prb{Phys. Rev. B }
\def\prc{Phys. Rev. C }
\def\prd{Phys. Rev. D }
\def\pre{Phys. Rev. E }
\def\prl{Phys. Rev. Lett. }
\def\prsa{Proc. R. Soc. A }
\def\psfg{Proteins: Struct., Func. and Gen. }
\def\sci{Science }
\def\spj{Sov. Phys. JETP }
\def\ss{Surf. Sci. }
\def\tca{Theor. Chim. Acta }
\def\tpa{Theor. Prob. and Appl. }
\def\zpb{Z. Phys. B. }
\def\zpc{Z. Phys. Chem. }
\def\zpd{Z. Phys. D }
\def\aciee{Angew. Chem. Int. Ed. Engl. }
\def\ac{Acta. Crystallogr. }
\def\acp{Adv. Chem. Phys. }
\def\acr{Acc. Chem. Res. }
\def\ajp{Am. J.~Phys. }
\def\ap{Adv. Phys. }
\def\arpc{Ann. Rev. Phys. Chem. }
\def\cccc{Coll. Czech. Chem. Comm. }
\def\cpl{Chem. Phys. Lett. }
\def\crev{Chem. Rev. }
\def\dalton{J.~Chem. Soc., Dalton Trans. }
\def\el{Europhys. Lett. }
\def\faraday{J.~Chem. Soc., Faraday Trans. }
\def\fartrans{J.~Chem. Soc., Faraday Trans. }
\def\fdisc{J.~Chem. Soc., Faraday Discuss. }
\def\ic{Inorg. Chem. }
\def\ijqc{Int. J.~Quant. Chem. }
\def\jcis{J.~Colloid Interface Sci. }
\def\jcsft{J.~Chem. Soc., Faraday Trans. }
\def\jacs{J.~Am. Chem. Soc. }
\def\jas{J.~Atmos. Sci. }
\def\jcc{J.~Comp. Chem. }
\def\jcp{J.~Chem. Phys. }
\def\jce{J.~Chem. Ed. }
\def\jcscc{J.~Chem. Soc., Chem. Commun. }
\def\jetp{J.~Exp. Theor. Phys. (Russia) }
\def\jmc{J.~Math. Chem. }
\def\jmp{J.~Math. Phys. }
\def\jmsp{J.~Mol. Spec. }
\def\jmst{J.~Mol. Structure }
\def\jncs{J.~Non-Cryst. Solids }
\def\jpc{J.~Phys. Chem. }
\def\jpcm{J.~Phys. Condensed Matter. }
\def\jpsj{J.~Phys. Soc. Jpn. }
\def\jsp{J.~Stat. Phys. }
\def\mg{Math. Gazette }
\def\mp{Mol. Phys. }
\def\mpns{Mol. Phys.}
\def\nat{Nature }
\def\pac{Pure. Appl. Chem. }
\def\phys{Physics }
\def\pla{Phys. Lett. A }
\def\plb{Phys. Lett. B }
\def\phm{Philos. Mag. }
\def\pmb{Philos. Mag. B }
\def\pnas{Proc.\ Natl.\ Acad.\ Sci.\  USA }
\def\pr{Phys. Rev. }
\def\pra{Phys. Rev. A }
\def\prb{Phys. Rev. B }
\def\prc{Phys. Rev. C }
\def\prd{Phys. Rev. D }
\def\pre{Phys. Rev. E }
\def\prl{Phys. Rev. Lett. }
\def\prsa{Proc. R. Soc. A }
\def\ss{Surf. Sci. }
\def\sci{Science }
\def\tca{Theor. Chim. Acta }
\def\zpc{Z. Phys. Chem. }
\def\zpd{Z. Phys. D }
\def\zfpd{Z. Phys. D }
\def\zpdamc{Z. Phys. D }
\def\aciee{Angew. Chem. Int. Ed. Engl. }
\def\ac{Acta. Crystallogr. }
\def\acp{Adv. Chem. Phys. }
\def\acr{Acc. Chem. Res. }
\def\ajp{Am. J.~Phys. }
\def\am{Adv. Mater. }
\def\apl{Appl. Phys. Lett. }
\def\arpc{Ann. Rev. Phys. Chem. }
\def\mrsb{Mater. Res. Soc. Bull. }
\def\cccc{Coll. Czech. Chem. Comm. }
\def\cj{Comput. J. }
\def\cp{Chem. Phys. }
\def\cpc{Comp. Phys. Comm. }
\def\cpl{Chem. Phys. Lett. }
\def\crev{Chem. Rev. }
\def\el{Europhys. Lett. }
\def\fd{Faraday Disc. }
\def\ic{Inorg. Chem. }
\def\ijmpc{Int. J.~Mod. Phys. C }
\def\ijqc{Int. J.~Quant. Chem. }
\def\jcis{J.~Colloid Interface Sci. }
\def\jcsft{J.~Chem. Soc., Faraday Trans. }
\def\jacs{J.~Am. Chem. Soc. }
\def\jap{J.~Appl. Phys. }
\def\jas{J.~Atmos. Sci. }
\def\jcc{J.~Comp. Chem. }
\def\jcp{J.~Chem. Phys. }
\def\jce{J.~Chem. Ed. }
\def\jcscc{J.~Chem. Soc., Chem. Commun. }
\def\jetp{J.~Exp. Theor. Phys. (Russia) }
\def\jmsp{J.~Mol. Spec. }
\def\jmst{J.~Mol. Structure }
\def\jncs{J.~Non-Cryst. Solids }
\def\jpa{J.~Phys. A }
\def\jpc{J.~Phys. Chem. }
\def\jpcssp{J.~Phys. C: Solid State Phys. }
\def\jpca{J.~Phys. Chem. A. }
\def\jpcb{J.~Phys. Chem. B. }
\def\jpcm{J.~Phys. Condensed Matter. }
\def\jpcs{J.~Phys. Chem. Solids. }
\def\jpsj{J.~Phys. Soc. Jpn. }
\def\jpfmp{J.~Phys. F, Metal Phys. }
\def\mg{Math. Gazette }
\def\msr{Mater. Sci. Rep. }
\def\nat{Nature }
\def\njc{New J.~Chem. }
\def\njp{New J.~Phys. }
\def\pac{Pure. Appl. Chem. }
\def\phys{Physics }
\def\pma{Philos. Mag. A }
\def\pmb{Philos. Mag. B }
\def\pml{Philos. Mag. Lett. }
\def\pnasu{Proc. Natl. Acad. Sci. USA }
\def\pr{Phys. Rev. }
\def\prep{Phys. Reports }
\def\pra{Phys. Rev. A }
\def\prb{Phys. Rev. B }
\def\prc{Phys. Rev. C }
\def\prd{Phys. Rev. D }
\def\pre{Phys. Rev. E }
\def\prl{Phys. Rev. Lett. }
\def\prsa{Proc. R. Soc. A }
\def\pss{Phys. State Solidi }
\def\pssb{Phys. State Solidi B }
\def\rmp{Rev. Mod. Phys. }
\def\rpp{Rep. Prog. Phys. }
\def\sci{Science }
\def\ss{Surf. Sci. }
\def\tca{Theor. Chim. Acta }
\def\tetra{Tetrahedron }
\def\tams{Trans. Am. Math. Soc. }
\def\zpb{Z. Phys. B. }
\def\zpc{Z. Phys. Chem. }
\def\zpd{Z. Phys. D }
\def\currentyear{2006}

\bibliographystyle{thesis}
\bibliography{trygub}
\clearpage

\section*{Tables}

\begin{table}[h]
\caption{\label{table:four}Results for \LJ{55} at $k_BT/\epsilon=0.4$ for $t_{\rm b}$ waiting
times averaged over 1000 KMC trajectories and a fractional convergence criterion
of $10^{-3}$ in the committor probability calculation. 
CPU times (seconds) are given in brackets for an UltraSparcIII 900\,MHz processor.
The two timings in the $k^{\rm NSS}_{\rm AB}$ column refer to runs with/without
leapfrog moves.
For comparison,
$k^{\rm KMC}_{\rm AB}=0.48\times10^{-7}$ averaged over 1000 trajectories,
and this calculation required about 4800\,s on an UltraSparcIII 900\,MHz processor
both with and without leapfrog KMC moves.
All rate constants are in reduced units of $\nuLJ$.}
\begin{center}
\begin{tabular}{ccccl}
\hline\hline
$E_{\rm th}/\epsilon$ & $k^{\rm SS}_{\rm AB}\times10^7$ & $k^{\rm NSS}_{\rm AB}\times10^7$ & 
               $k^{\rm NSS}_{\rm AB}/k^{\rm SS}_{\rm AB}$ & $k^{\rm GT}_{\rm AB}\times10^7$ \\
\hline
$-266$ & $1.02$\,(1) & $0.50$\,(8/10) & 0.50 & 0.50\,(7) \\
$-267$ & $0.46$\,(1) & $0.41$\,(7/9) & 0.89 & 0.34\,(8) \\
$-268$ & $0.46$\,(1) & $0.41$\,(6/6) & 0.90 & 0.34\,(9) \\
$-269$ & $0.46$\,(1) & $0.45$\,(5/5) & 0.96 & 0.38\,(8) \\
$-270$ & $0.46$\,(1) & $0.44$\,(4/5) & 0.95 & 0.38\,(9) \\
$-271$ & $0.53$\,(1) & $0.16$\,(11/8) & 0.29 & 0.14\,(22) \\
$-272$ & $0.54$\,(3) & $0.15$\,(6/5) & 0.28 & 0.15\,(19) \\
$-273$ & $0.58$\,(3) & $0.16$\,(5/5) & 0.27 & 0.15\,(22) \\
$-274$ & $1.20$\,(3) & $0.23$\,(4/4) & 0.19 & 0.24\,(25) \\
$-275$ & $1.20$\,(3) & $0.23$\,(4/4) & 0.19 & 0.24\,(25) \\
$\le -276$ & $2.27$\,(3) & $0.42$\,(4/3) & 0.18 & 0.42\,(24) \\
\hline\hline
\end{tabular}
\end{center}
\end{table}

\newpage

\begin{table}[h]
\caption{\label{table:three}Results for \LJ{55} at $k_BT/\epsilon=0.3$ for $t_{\rm b}$ waiting
times averaged over 1000 KMC trajectories, and fractional convergence criteria 
in the committor probability calculation
of $10^{-3}$ for $E_{\rm th}\ge-270\,\epsilon$
and $10^{-7}$ for $E_{\rm th}\le-271\,\epsilon$.
CPU times (seconds) are given in brackets for an UltraSparcIII 900\,MHz processor.
The two timings in the $k^{\rm NSS}_{\rm AB}$ column refer to runs with/without
leapfrog moves.
For comparison,
$k^{\rm KMC}_{\rm AB}=0.13\times10^{-11}$ averaged over 1000 trajectories,
and this calculation required about $2.0\times10^7$\,s and $2.6\times10^7$\,s
on an UltraSparcIII 900\,MHz processor for KMC runs
with and without leapfrog KMC moves, respectively.
All rate constants are in reduced units of $\nuLJ$.}
\begin{center}
\begin{tabular}{cllcl}
\hline\hline
$E_{\rm th}/\epsilon$ & $k^{\rm SS}_{\rm AB}\times10^{11}$ & $k^{\rm NSS}_{\rm AB}\times10^{11}$ & 
               $k^{\rm NSS}_{\rm AB}/k^{\rm SS}_{\rm AB}$ & $k^{\rm GT}_{\rm AB}\times10^{11}$ \\
\hline
$-266$ & $0.19\,(2)$ & $0.16\,(3/4)$ & 0.85 & 0.16\,(7) \\
$-267$ & $0.13\,(1)$ & $0.12\,(2/2)$ & 0.96 & 0.10\,(8) \\
$-268$ & $0.13\,(1)$ & $0.12\,(2/2)$ & 0.96 & 0.10\,(9) \\
$-269$ & $0.12\,(1)$ & $0.12\,(1/2)$ & 1.00 & 0.11\,(8) \\
$-270$ & $0.12\,(1)$ & $0.12\,(1/2)$ & 0.99 & 0.10\,(9) \\
$-271$ & $0.13\,(122)$ & $0.11\,(125/135)$ & 0.84 & 0.038\,(22) \\
$-272$ & $0.13\,(128)$ & $0.11\,(126/129)$ & 0.83 & 0.038\,(19) \\
$-273$ & $0.13\,(128)$ & $0.12\,(129/129)$ & 0.87 & 0.040\,(22) \\
$-274$ & $0.16\,(146)$ & $0.10\,(147/147)$ & 0.61 & 0.067\,(25) \\
$-275$ & $0.16\,(147)$ & $0.10\,(147/147)$ & 0.61 & 0.067\,(25) \\
$\le -276$ & $0.17\,(152)$ & $0.13\,(152/152)$ & 0.77 & 0.10\,(24) \\
\hline\hline
\end{tabular}
\end{center}
\end{table}

\newpage

\begin{table}[h]
\caption{\label{table:two}Results for \LJ{55} at $k_BT/\epsilon=0.2$ for $t_{\rm b}$ waiting
times averaged over 1000 KMC trajectories, and fractional convergence criteria 
in the committor probability calculation
of $10^{-3}$ for $E_{\rm th}\ge-268\,\epsilon$,
$10^{-4}$ for $-270\,\epsilon \le E_{\rm th}\le-269\,\epsilon$,
and $10^{-9}$ for $E_{\rm th}=-271\,\epsilon$.
The $k^{\rm SS}_{\rm AB}$ result for
$E_{\rm th}=-271\,\epsilon$ is probably not converged even for this
tighter convergence criterion, and should be considered a lower bound
for comparison with $k^{\rm GT}_{\rm AB}$.
CPU times (seconds) are given in brackets for an UltraSparcIII 900\,MHz processor.
The two timings in the $k^{\rm NSS}_{\rm AB}$ column refer to runs with/without
leapfrog moves.
Conventional KMC calculations were not feasible at this temperature.
All rate constants are in reduced units of $\nuLJ$.}
\begin{center}
\begin{tabular}{cllcl}
\hline\hline
$E_{\rm th}/\epsilon$ & $k^{\rm SS}_{\rm AB}\times10^{22}$ & $k^{\rm NSS}_{\rm AB}\times10^{22}$ & 
               $k^{\rm NSS}_{\rm AB}/k^{\rm SS}_{\rm AB}$ & $k^{\rm GT}_{\rm AB}\times10^{22}$ \\
\hline
$-266$ & $0.79\,(2)$ & $0.74\,(2/635)$ & 0.93 & 0.75\,(9) \\
$-267$ & $0.65\,(1)$ & $0.65\,(2/36)$ & 1.00 & 0.65\,(10)\\
$-268$ & $0.65\,(1)$ & $0.65\,(2/28)$ & 1.00 & 0.65\,(10)\\
$-269$ & $0.65\,(1)$ & $0.65\,(2/219)$ & 1.00 & 0.65\,(10)\\
$-270$ & $0.65\,(1)$ & $0.65\,(2/206)$ & 1.00 & 0.65\,(10)\\
$-271$ & $0.092\,(119331)$ & --- & --- & 0.021\,(26)\\
$-272$ &--- & --- & --- & 0.022\,(23)\\
$-273$ &--- & --- & --- & 0.023\,(27)\\
$-274$ &--- & --- & --- & 0.044\,(31)\\
$-275$ &--- & --- & --- & 0.045\,(31)\\
$\le -276$ & --- & --- & --- & 0.60\,(29)\\
\hline\hline
\end{tabular}
\end{center}
\end{table}

\newpage

\begin{table}[h]
\caption{\label{table:one}Results for \LJ{55} at $k_BT/\epsilon=0.1$ for $t_{\rm b}$ waiting
times averaged over 1000 KMC trajectories, and a fractional convergence criterion
in the committor probability calculation
of $10^{-3}$ for $E_{\rm th}\ge-270\,\epsilon$.
For thresholds of $E_{\rm th}\le-271\,\epsilon$ it proved impossible to converge
the committor probability calculation for the fractional convergence criteria
required to yield meaningful results.
CPU times (seconds) are given in brackets for an UltraSparcIII 900\,MHz processor.
The timings in the $k^{\rm NSS}_{\rm AB}$ column refer to runs with
leapfrog moves; the corresponding runs without leapfrog moves were not feasible.
Conventional KMC calculations were also unfeasible at this temperature.
All rate constants are in reduced units of $\nuLJ$.}
\begin{center}
\begin{tabular}{cllcl}
\hline\hline
$E_{\rm th}/\epsilon$ & $k^{\rm SS}_{\rm AB}\times10^{52}$ & $k^{\rm NSS}_{\rm AB}\times10^{52}$ & 
               $k^{\rm NSS}_{\rm AB}/k^{\rm SS}_{\rm AB}$ & $k^{\rm GT}_{\rm AB}\times10^{52}$ \\
\hline
$-266$ & $0.14\,(2)$ & $0.12\,(3)$ & 0.87 & 0.12\,(8)\\
$-267$ & $0.14\,(2)$ & $0.14\,(544)$ & 1.00 & 0.14\,(9)\\
$-268$ & $0.14\,(2)$ & $0.14\,(3)$ & 1.00 & 0.14\,(11)\\
$-269$ & $0.14\,(3)$ & $0.14\,(3)$ & 1.00 & 0.14\,(10) \\
$-270$ & $0.14\,(3)$ & $0.14\,(20)$ & 1.00 & 0.14\,(10) \\
$-271$ & --- & --- & --- &  0.012\,(26) \\
$-272$ &--- & --- & --- & 0.012\,(23)\\
$-273$ &--- & --- & --- & 0.013\,(25)\\
$-274$ &--- & --- & --- & 0.075\,(25)\\
$-275$ &--- & --- & --- & 0.075\,(25)\\
$\le -276$ & --- & --- & --- & 0.13\,(24)\\
\hline\hline
\end{tabular}
\end{center}
\end{table}

\clearpage

\begin{figure}[htb]
\psfrag{V}[Bl][Bl]{\fsz$V/\epsilon$}
\psfrag{-264}[Br][Br]{\fsz$-264$}
\psfrag{-265}[Br][Br]{\fsz$-265$}
\psfrag{-266}[Br][Br]{\fsz$-266$}
\psfrag{-267}[Br][Br]{\fsz$-267$}
\psfrag{-268}[Br][Br]{\fsz$-268$}
\psfrag{-269}[Br][Br]{\fsz$-269$}
\psfrag{-270}[Br][Br]{\fsz$-270$}
\psfrag{-271}[Br][Br]{\fsz$-271$}
\psfrag{-272}[Br][Br]{\fsz$-272$}
\psfrag{-273}[Br][Br]{\fsz$-273$}
\psfrag{-274}[Br][Br]{\fsz$-274$}
\psfrag{-275}[Br][Br]{\fsz$-275$}
\psfrag{-276}[Br][Br]{\fsz$-276$}
\psfrag{-277}[Br][Br]{\fsz$-277$}
\psfrag{-278}[Br][Br]{\fsz$-278$}
\psfrag{-279}[Br][Br]{\fsz$-279$}
\psfrag{Ih}[cc][cc]{\fsz$ I_h$}
\psfrag{C5v}[cc][cc]{\fsz$ C_{5v}$}
\psfrag{C2v}[cc][cc]{\fsz$ C_{2v}$}
\psfrag{A}[cc][cc]{\fsz A}
\psfrag{B}[cc][cc]{\fsz B}
\centerline{\includegraphics[width=0.80\textwidth]{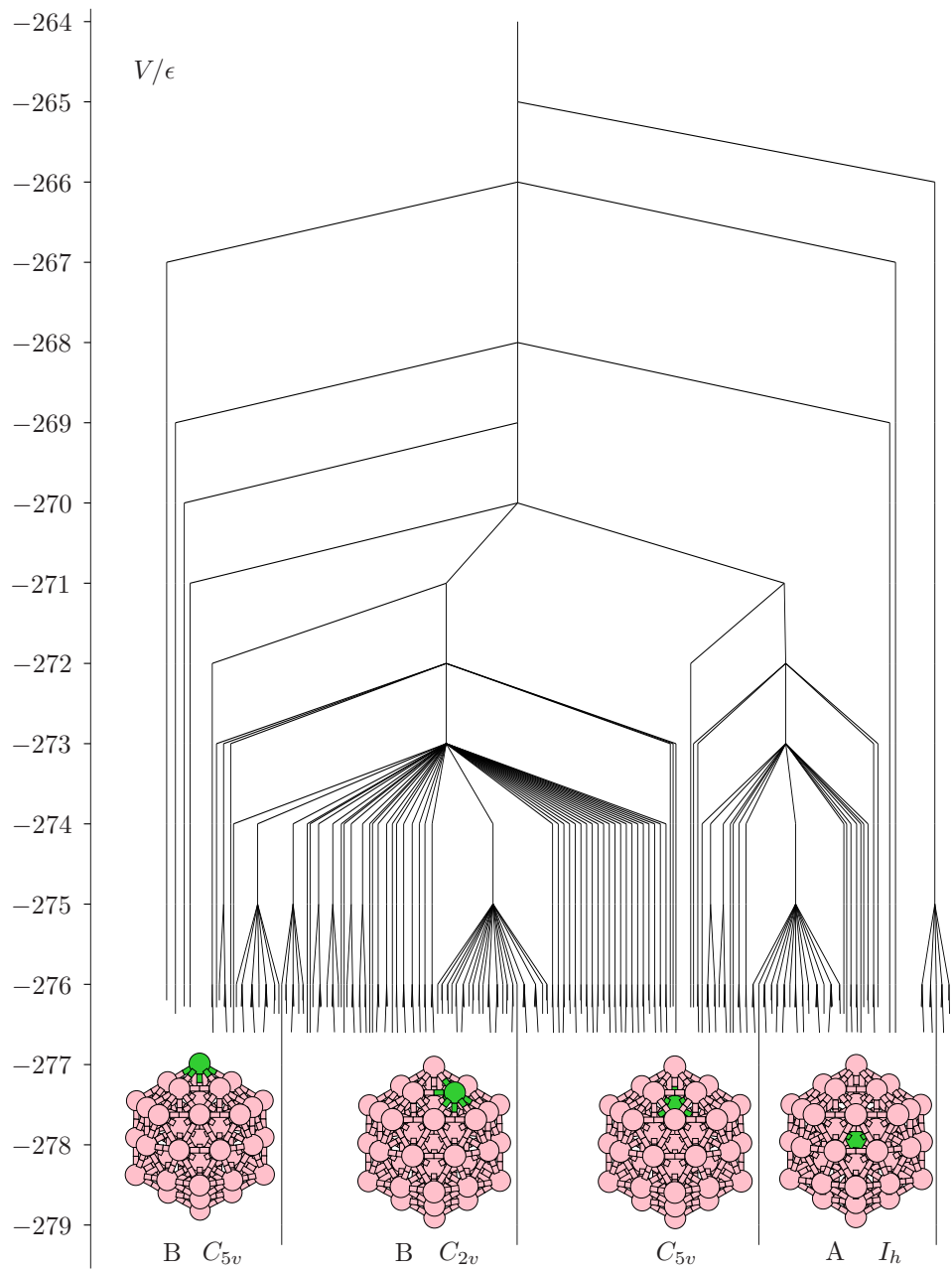}}
\caption{\label{fig:LJ55.permtree}
Disconnectivity graph for \LJ{55}
in which permutation-inversion isomers of the shaded atom are distinguished. 
Permutation-inversion isomers
of the other atoms are grouped together for every minimum and transition state.
The tagged atom can occupy four distinct sites in the icosahedral global 
minimum, producing four separate branches,
which are labelled according to the appropriate point
group symmetry and A/B region (before any reclassification).
The DPS stationary point database contains 5,609 minima and 10,134 transition states \cite{Wales02},
but the graph includes only the lowest 250 minima for clarity.
The potential energy, $V$, is in units of $\epsilon$.
}
\end{figure}

\end{document}